\documentclass{article}
\usepackage[nomove]{overcite}
\usepackage{epsfig}

\def\thebibliography#1{\section*{References\markboth
{REFERENCES}{REFERENCES}}\list
{\arabic{enumi}. }{\settowidth\labelwidth{[#1]}\leftmargin\labelwidth
\advance\leftmargin\labelsep
\usecounter{enumi}}
\def\newblock{\hskip .11em plus .33em minus -.07em}
\sloppy
\sfcode`\.=1000\relax}

\begin{document}
\title{\vspace*{-1.6cm}
Geometry dominated fluid adsorption on sculptured substrates.}
\author{C.\ Rasc\'{o}n, A.\ O.\ Parry\\
Department of Mathematics, Imperial College\\ 180 Queen's Gate,
London SW7 2BZ, United Kingdom}
\date{}
\maketitle

{\bf
Experimental methods allow the shape\cite{-1,-3} and chemical
composition\cite{-2} of solid surfaces to be controlled at a
mesoscopic level. Exposing such structured substrates to
a gas close to coexistence with its liquid can produce
quite distinct adsorption characteristics compared to that
occuring for planar systems\cite{Rough}, which may well play an 
important role in developing technologies such as
super-repellent surfaces\cite{Rep1,Rep2} or
micro-fluidics\cite{MicF1,MicF2}. Recent studies 
have concentrated on adsorption of liquids at
rough\cite{Kardar,Stella,Netz} and 
heterogeneous\cite{MicF3} substrates and the
characterisation of nanoscopic liquid films\cite{Mon}.
However, the fundamental effect of geometry 
has hardly been addressed.
Here we show that varying the shape of the substrate can exert a
profound influence on the adsorption isotherms allowing
us to smoothly connect wetting and capillary condensation through
a number of novel and distinct examples of fluid interfacial phenomena.
This opens the possibility of tailoring the adsorption
properties of solid substrates by sculpturing their surface
shape.}

The interaction of a gas with a solid surface is usually mediated by an  
adsorbed (microscopically thin) layer of liquid which is ultimately  
responsible for the rich behaviour of the solid-gas interface\cite{1}.
When the gas coexists with its liquid, an increase in the temperature is accompanied  
by an increase in the thickness of this adsorbed layer which attains  
macroscopic character at the wetting temperature $T_w$. At this  
temperature, the contact angle of a macroscopic drop placed on the surface  
vanishes and the liquid, which is said to wet that solid surface,
spreads. Alternatively, this situation can be reached for a fixed
temperature $T$  
above $T_w$ (but below the critical temperature $T_c$) by increasing the  
pressure towards the coexistence value $p_{co}(T)$ or, equivalently,
by increasing the chemical potential towards $\mu_{co}(T)$ (see Fig.\ 1).
This phase transition, known as {\em complete wetting}, has been intensely  
studied for fundamental as well  
as for important technological reasons\cite{1}.
Theoretical predictions confirmed by  
experiments have shown that the thickness $\ell_{\pi}$ of the adsorbed layer
diverges when it approaches coexistence with a non-universal exponent
$\beta_{co}$ which is dependent on the range of the
(solid-fluid and fluid-fluid) particle interactions involved in the system:
$\ell_{\pi}\!\sim\!|\Delta\mu|^{-\beta_{co}}$, where 
$\Delta\mu\!\equiv\!\mu\!-\!\mu_{co}$ is the chemical potential
(relative to the chemical potential at coexistence)\cite{1}
which, for dilute gases, is related to the partial pressure by
$\Delta\mu\!=\!k_{B}T\log(p/p_{co}(T))$ where $k_B$ is
Boltzmann's constant.
This prediction, however, is restricted to planar solid substrates. 

One of the simplest geometrical configurations beyond the planar wall
is a parallel slit geometry comprising two planar substrates a distance $\cal L$ from each
other. For this system, a different phenomenon occurs called
{\em capillary condensation}\cite{Chris,3}. By increasing the chemical potential
at a constant temperature $T\!>\!T_w$,
the thickness of the adsorbed liquid layers
on either wall grows according to a similar law as in the planar system.
However, when the chemical potential reaches a certain value
$\Delta\mu^{*}\!<\!0$, the space between the walls fills up with
liquid, even though that phase is {\em not} thermodynamically stable
in bulk. The location of this first-order phase transition is
given, for sufficiently large values of $\cal L$, by the Kelvin
equation,
\begin{equation}
\Delta\mu^{*}=-\frac{2\,\sigma}{(\rho_L-\rho_G)\,{\cal L}},
\end{equation}
where $\rho_L$ and $\rho_G$ are the densities of the liquid
and the gas respectively and $\sigma$ is the surface tension between
these phases.

This simple comparison reveals the striking influence of geometry
on adsorption isotherms: the continuous divergence dominated by
the microscopic details of the interaction with the substrate
(embodied in the exponent $\beta_{co}$), in the case of a planar wall, is
transformed into a discontinuous jump at a value of the chemical
potential $\Delta\mu^{*}$, which does not depend on microscopic
properties of the substrate, in the case of the slit (see Fig.\ 1).
Interestingly, whilst the full phenomenology of
wetting is understood with the help of effective interfacial
models\cite{1}, capillarity condensation is best described by density
functional methods\cite{3}.\\

Here we show, with a purely geometrical model, that there
is a wealth of adsorption phenomena induced by different wall
geometries, which smoothly interpolates between
wetting and capillary condensation.
The model, exact in the macroscopic limit,
does not intend to be a microscopic description of these phenomena
but to capture the essential features of geometrically
dominated adsorption, which, due to the lack of adequate models,
remains largely unexplored. 

In order to understand the subtle connection between
wetting and capillarity,
we consider generalised solid wedges with cross section
described by a shape function $|x|^\gamma/L^{\gamma\!-\!1}$ in the $x$
direction (where $L$ is a length associated with the dimensions of
the wall).
The limiting cases $\gamma\!=\!0$ and $\gamma\!=\!\infty$,
correspond to a planar substrate and a capillary
slit respectively. Other notable intermediate cases 
include the linear ($\gamma\!=\!1$) and parabolic ($\gamma\!=\!2$)
wedges as shown in Fig.\ 2. This system has been only
partially studied: macroscopic\cite{Finn,Pomeau,Hauge} and
effective interfacial methods have 
been used to describe the adsorption isotherms at
a linear wedge\cite{Neimark,4,5} and at walls with $0\!<\!\gamma\!\leq\!1$,
(in the asymptotic limit $\Delta\mu\!\rightarrow\!0$)\cite{6}.
Although for $\gamma\!>\!1$ effective interfacial methods
are problematic, 
they suggest a picture of isothermal adsorption dominated
by geometry, thereby supporting the following geometrical approach.
For a given temperature and chemical potential, we can
determine two pertinent lengths: the thickness of an adsorbed
layer on a planar substrate, $\ell_\pi$, and the radius of
curvature given by the Laplace equation
$R\!\equiv\!\sigma/(\rho_L\!-\!\rho_G)|\Delta\mu|$.
In order to construct the equilibrium profile of the adsorbed layer,
we first {\em coat} the surface with a
layer of thickness $\ell_\pi$ (Fig.\ 3(b)).
This is necessary to allow, in certain regimes, for the influence
of microscopic forces (which are not accounted for
in a purely macroscopic approach\cite{Hauge}).
After that, a cylinder of radius $R$ is fitted at
the point of maximum curvature of the coated surface (Fig.\ 3(c)).
The tangency points are the edges of a
meniscus, whose shape is the lower part of the cylinder.
The resulting interfacial shape can be determined by the
combination of the coated surface and the meniscus (if any),
and allow us to calculate the thickness $\ell_0$ of the adsorbed
layer at the midpoint ($x\!=\!0$) or the size of the meniscus
as we approach coexistence (Fig.\ 3(d)).

This deceptively simple geometrical construction describes
the gradual transformation of wetting (dominated by
intermolecular forces) into capillary condensation (essentially
dominated by geometry) as the substrate varies from being
flat to being a parallel slit.
The transformation is intricate and involves several
asymptotic and pre-asymptotic regimes classified
by the different behaviour of the midpoint thickness $\ell_0$
as a function of the chemical potential (Fig.\ 2).
In our calculations, we have considered the experimentally
relevant case of dispersion forces ($\beta_{co}\!=\!1/3$).
Besides, as we are interested in substrates with mesoscopic
structure for which the influence of gravity is
negligible, we have fixed the value of
$A/12\pi\sigma\,L^2\!\sim\!10^{-3}$,
where $A$ is the Hamaker constant of the planar system\cite{8}.
Considerably different values of this dimensionless
quantity modify the presence of pre-asymptotic regimes 
but do not introduce any new feature in the phase
diagram.

In the asymptotic limit ($\Delta\mu\!\rightarrow\!0$),
the continuous divergence of $\ell_0$ 
falls within one of three regimes, depending on the
shape of the wall: a planar regime (PL) for
$\gamma\!<\!\gamma^{*}\!\equiv\!2\beta_{co}/(1\!+\!\beta_{co})$,
for which $\ell_0\!\approx\!\ell_\pi\!\sim\!|\Delta\mu|^{-\beta_{co}}$,
a first geometrically dominated regime (G$_1$) for
$\gamma^{*}\!<\!\gamma\!<\!1$ for which 
$\ell_0\!\sim\!|\Delta\mu|^{-\gamma/(2\!-\!\gamma)}$
and a second geometrically dominated regime (G$_2$) for
$1\!<\!\gamma$ for which 
$\ell_0\!\sim\!|\Delta\mu|^{-\gamma}$.
Note that the exponent characterising the divergence of $\ell_0$
varies continuously with the wedge-geometry exponent $\gamma$.
The regimes PL and G$_1$ and 
the existence of a marginal case $\gamma\!=\!\gamma^{*}$ are
in perfect agreement with the aforementioned prediction
based on an effective Hamiltonian model\cite{6} thereby corroborating
the strong influence of geometry on liquid adsorption.
The regime
G$_2$, not previously described in the literature, can be employed
to fabricate microfluidic devices with an arbitrarily
sensitive response to changes in partial pressure.

Additionally, several pre-asymptotic regimes are also captured
by this description. Figure 4 shows the typical evolution
of the midpoint thickness $\ell_0$ as a function of the
chemical potential for different wall shapes.
For $\gamma\!=\!0.25$ ($\gamma\!<\!\gamma^*$),
two clear regimes can be distinguished (Fig.\ 4(a)). Together with the
planar asymptotic regime close to coexistence, a regime (GP)
involving geometric and planar features appears for which the
midpoint thickness grows like $\ell_0\!\sim\!\ell_{\pi}^\gamma$.
This regime persists for $\gamma^*\!<\!\gamma\!<\!1$, as can be
seen in Fig.\ 4(b) for $\gamma\!=\!0.75$,
although the asymptotic behaviour has changed and is geometrically
driven.
In contrast, for $\gamma\!>\!1$, along with a unique asymptotic
behaviour, there are two distinct pre-asymptotic regimes
separated by the marginal case $\gamma\!=\!2$. For
$1\!<\!\gamma\!<\!2$, $\ell_0$ grows in a planar-like manner
far from coexistence and presents a crossover to the
true asymptotic regime G$_2$ as can be seen in Fig.\ 4(c)
for  $\gamma\!=\!1.5$. Within our approximation, the chemical
potential always enters as the dimensionless variable
$\Delta\mu(\rho_L\!-\!\rho_G)L/\sigma$. Depending on
the value of each of these quantities, the experimental window
might happen to reach only one of the regimes described above.

The singular case $\gamma\!=\!2$, the parabolic wedge, deserves
especial attention since it gives rise to a new phenomenon. 
For this case, no meniscus forms until
the chemical potential has exceeded a certain threshold. This
emerges as a direct consequence of the geometry. If the
radius of curvature given by Laplace's equation is much smaller
than the minimum radius of curvature of the surface
($R\!\ll\!L/2$), no meniscus
is developed and a thin layer covers the substrate.
The value of the chemical potential for which the meniscus
starts to form is given implicitly, in this simple geometrical
approximation, by $R\!+\!\ell_{\pi}\!=\!L/2$,
where $\ell_{\pi}$ represents a correction due to the thickness
of the adsorbed layer. Once the meniscus has originated,
the midpoint thickness tends to grow as dictated by the
asymptotic limit G$_2$. We refer to this phenomenon as
the {\it meniscus transition} although fluctuation
effects, not described here, transform it into a smooth,
but nonetheless abrupt, crossover from microscopic to
macroscopic adsorption, characterised by scaling behaviour.
It can be consider a
legitimate phase transition in the macroscopic limit
($\ell_{\pi}/L\!\rightarrow\!0$) for which
the present geometrical model is exact. 

A new phenomenon, leading to capillary condensation in the limit
$\gamma\!\rightarrow\!\infty$, appears for $\gamma\!>\!2$.
For these substrates, the point of minimum curvature has
shifted from $x\!=\!0$ and, due to the symmetry
$x\!\leftrightarrow\!-x$, {\it two}
menisci materialise in two (symmetric) points of the wall
at a certain value of the chemical potential.
Since those two points are locally parabolic,
this effect is the same as the {\it meniscus transition}
described for the case $\gamma\!=\!2$. However, by
approaching coexistence further, these menisci extend in
size and eventually merge into a unique central meniscus.
This profoundly affects the behaviour of the system.
Once this unique meniscus is developed, the thickness at the
midpoint is extremely sensitive to changes in the
chemical potential and approximates the asymptotic
limit $\ell_0\!\sim\!|\Delta\mu|^{-\gamma}$.
In the limiting case $\gamma\!=\!\infty$,
the merging of the two menisci precipitates
the immediate and abrupt rise of the liquid level and
the space between the walls fills up completely,
giving rise to capillary condensation. Since the
parallel walls are connected by the bottom of the
wedge, no metastable states are expected\cite{9}.
The geometrical description of this phenomenon
predicts that the transitions takes place at a value
of the chemical potential given by
$R+\ell_{\pi}\!=\!L$, which recovers the
Kelvin equation (1) with a correction, given by $\ell_\pi$,
exact for systems with short-ranged intermolecular forces\cite{3}.

This extraordinarily rich sequence of different
regimes and phenomena
induced by the substrate shape
are shown in the phase diagram of Fig.\ 2.
A similar phase diagram is expected for radially
symmetric substrates described by a shape function
$\sim\!r^\gamma$, which range between a flat surface
and a capillary tube.
These distinct substrate shapes could be used as microfluidic
devices to place mesoscopic amounts of liquid in preferential
points of the surface of a solid substrate in a controlled
manner by simply varying the partial pressure. \\

\vspace{1cm}
{\bf Acknowledgements}\\
CR acknowledges economical support from the European Commission.

\begin{figure}
	\centerline{\epsfig{file=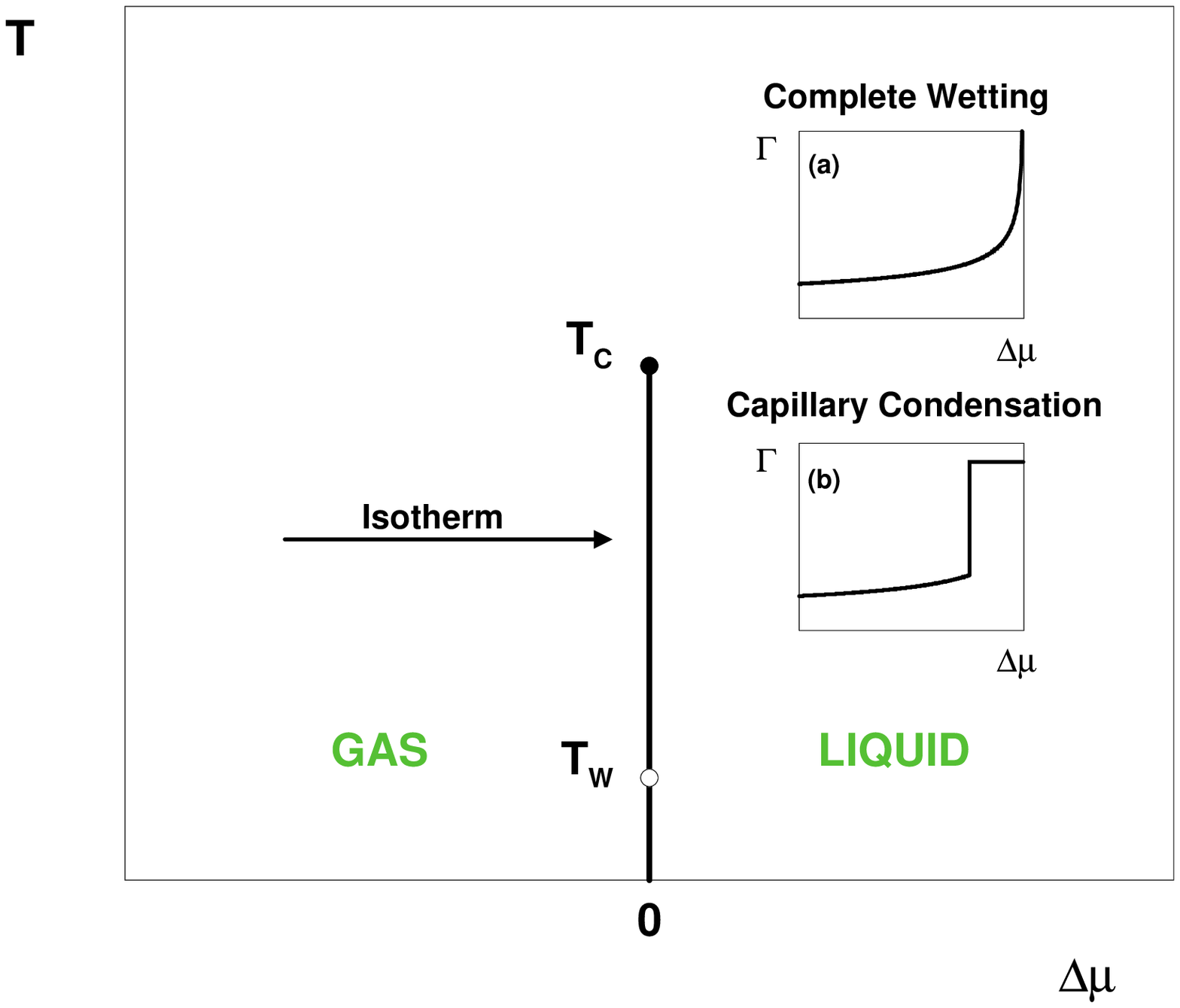,width=9.cm}}
	\caption{Schematic bulk phase diagram of a fluid showing the
the coexistence between gas and liquid phases. The gas phase is placed
in contact with a solid substrate at a temperature between the critical
temperature $T_c$ and the wetting temperature $T_w$. As we approach
coexistence, a liquid layer of increasing thickness is adsorbed
on the substrate. Adsorption isotherms are sketched for a
planar substrate (a) and for a parallel slit (b).}
	\label{One}
\end{figure}

\begin{figure}
	\centerline{\epsfig{file=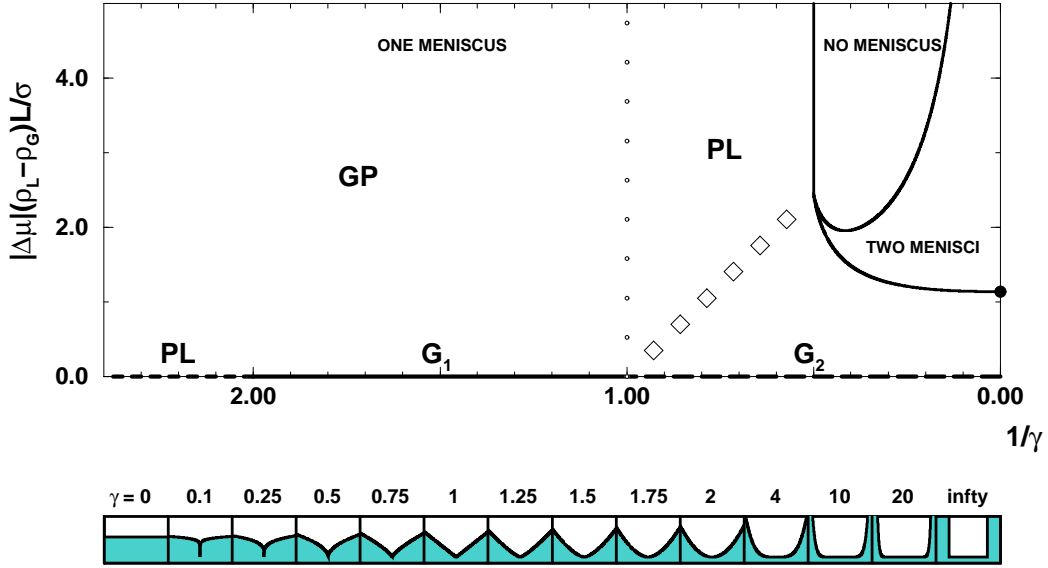,width=14.cm}}
\vspace*{-2cm}
	\caption{Phase diagram for the adsorption isotherms at
generalised solid wedges with cross section $\sim\!|x|^\gamma$
in the $x$ direction, as obtained using the geometrical
construction of Fig.\ 3. The thickness ($\ell_0$) of the adsorbed layer
at the mid-point ($x\!=\!0$) grows following different power laws
as a function of the chemical potential $\Delta\mu$. 
We have chosen the experimentally relevant case of dispersion
forces ($\beta_{0}\!=\!1/3\,\Rightarrow\,\gamma^{*}\!=\!1/2$)
and fixed the dimensionless variable $A/12\pi\sigma\,L^2\!\sim\!10^{-3}$,
where $A$ is the Hamaker constant of the planar system.
Approaching the bulk liquid-vapor coexistence, there
are three asymptotic regimes: planar (PL) for
$\gamma\!<\!\gamma^{*}\!\equiv\!2\beta_{co}/(1\!+\!\beta_{co})$,
and two geometrically dominated regimes G$_1$ and G$_2$.
Furthermore, different pre-asymptotic regimes appear for
larger values of $|\Delta\mu|$: for $\gamma\!<\!1$, a regime
combining planar and geometric features (GP), a planar regime
for $1\!<\!\gamma\!<\!2$, a regime in which no meniscus is
formed, and a regime with two symmetrically placed meniscus.
In the limit $\gamma\!\rightarrow\!\infty$, the transition
between the two menisci regime and the central meniscus
regime gives rise to capillary condensation (filled dot).
The lower diagram shows the evolution of the cross section
with the parameter $\gamma$ ranging from a planar wall ($\gamma\!=\!0$)
to a capillary slit ($\gamma\!=\!\infty$).}
	\label{Two}
\end{figure}

\begin{figure}
	\centerline{\epsfig{file=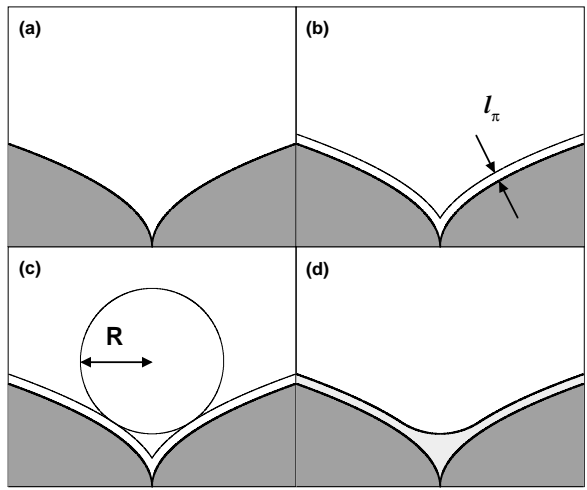,width=9.cm}}
	\caption{Geometrical construction for determining the shape of
the adsorbed layer. The surface (a) is first {\it coated} with a thin film of
thickness $\ell_\pi$, the thickness of an adsorbed layer on a planar
susbtrate (b). Then, a cylinder of radius $R$, given by Lapace's equation,
is fitted in the point of minimum curvature (c). The final shape is
given by the continous combination of both shapes (d).}
	\label{Three}
\end{figure}

\begin{figure}
	\centerline{\epsfig{file=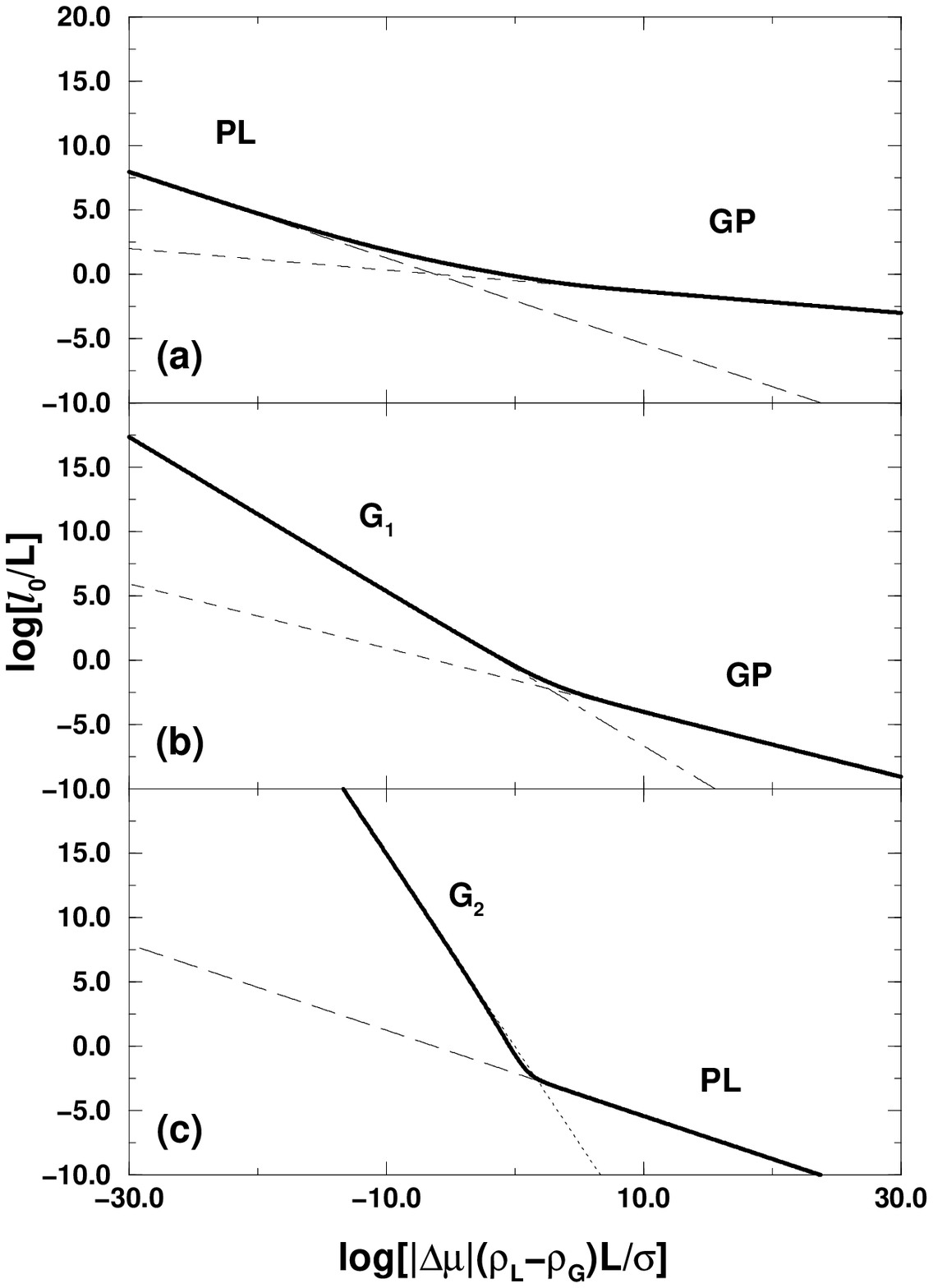,width=12.cm}}
	\caption{Adsorption isotherms for solid wedges with cross
section described by a shape function $\sim\!|x|^\gamma$ in the $x$
direction. The thickness of the adsorbed layer at the
midpoint $x\!=\!0$, calculated with the geometrical method
described in the text, is represented as a function of the chemical
potential $\Delta\mu$ for (a) $\gamma\!=\!0.25$, (b) $\gamma\!=\!0.75$
and (c) $\gamma\!=\!1.5$. The scale is the same for all the wall shapes
in order to facilitate the comparison. The pure asymptotic regimes
correspond to the straight lines: PL (long dashed), GP (short dashed),
G$_1$ (dot-dashed) and G$_2$ (dotted).}
	\label{Four}
\end{figure}


\begin{thebibliography}{99}
\addcontentsline{toc}{section}{References}
\bibitem{-1} Xia, Y.\ \& Whitesides, G.M.\ Soft Lithography.
{\it Angew.\ Chem.\ Int.\ Ed.\ } {\bf 37} 550-575 (1998).
\bibitem{-3} Trau, M., Yao, N., Kim, E., Xia, Y., Whitesides, G.M.\
\& Aksay, I.A.\ Microscopic patterning of orientated mesoscopic
silica through guided growth. {\it Nature}, {\bf 390}, 674-676 (1997).
\bibitem{-2} Kumar, A., Abbott, N.A., Kim, E., Biebuyck, H.A.\ \&
Whitesides, G.M.\ Patterned self-assembled monolayers and
meso-scale phenomena. {\it Acc.\ Chem.\ Res.\ } {\bf 28}, 219-226 (1995).
\bibitem{Rough} Dietrich, S.\ in Proceedings of the NATO-ASI
"New Approaches to Old and New Problems in
Liquid State Theory" (1998)
(ed., C.\ Caccamo, J.P.\ Hansen, G.\ Stell).
\bibitem{Rep1} Shibuichi, S., Yamamoto, T., Onda T.\ \&
Tsujii, K.\ Super water- and oil-repellent surfaces resulting
from fractal structure. {\it J.\ Colloid Interface
Sci.\ } {\bf 208}, 287-294 (1998).
\bibitem{Rep2} Bico, J., Marzolin, C.\ \& Quere, D.\
Pearl drops. {\it Europhys.\ Lett.\ } {\bf 47}, 220-226 (1999).
\bibitem{MicF1} Weigl, B.H. \& Yager, P.\ Microfluidics -
Microfluidic diffusion-based separation and detection.
{\it Science} {\bf 283}, 346-347 (1999).
\bibitem{MicF2} Gravensen, P., Branebjerg, J.\ \& Jensen, O.S.\
Microfluidics - a review. {\it J.\ Micromech.\ Microeng.\ } {\bf 3},
168-182 (1993).
\bibitem{Kardar} Kardar, M.\ \& Indekeu, J.O.\ Adsorption
and wetting transitions on rough substrates. {\it Europhys.\ Lett.\ }
{\bf 12}, 161-166 (1990).
\bibitem{Stella} Giugliarelli, G.\ \& Stella,  A.L.\ 
Discontinuous interface depinning from a rough wall.
{\it Phys.\ Rev.\ E}{\bf 53}, 5035-5038 (1996).
\bibitem{Netz} Netz, R.R.\ \& Andelman, D.\
Roughness-induced wetting. {\it Phys.\ Rev.\ E}{\bf 55},
687-700 (1997).
\bibitem{MicF3} Gau, H., Herminghaus, S., Lenz, P.\ \& Lipowsky, R.\
Liquid morphologies on structured surfaces: From microchannels to microchips.
{\it Science} {\bf 283}, 46-49 (1999).
\bibitem{Mon} Luna, M., Colchero, J.\ \& Baro, A.M.\ Study of water
droplets and films on graphite by noncontact scanning force microscopy.
{\it J.\ Phys.\ Chem.\ } B {\bf 103} 9576-9581 (1999).
\bibitem{1} Dietrich, S.\ Wetting phenomena in
{\it Phase Transitions and Critical Phenomena},
(C.\ Domb and J.L.\ Lebowitz, eds.), Vol.\ {\bf 12}, 1-218
(Academic Press, London, 1988).
\bibitem{3} Evans, R., Marconi, U.M.B.\ \& Tarazona, P.\ Fluids in narrow
pores: Adsorption, capillary condensation, and critical points.
{\it  J.\ Chem.\ Phys.\ } {\bf 84}, 2376 (1986).
\bibitem{Chris} Christenson, H.K.\ Capillary condensation due to
van der Waals atraction in wet slits.
{\it Phys.\ Rev.\ Lett.\ } {\bf 73}, 1821-1824 (1994).
\bibitem{Finn} Concus, P.\ \& Finn, R.\ On the behaviour
of a capillary surface in a wedge. {\it Proc.\ Nat.\ Acad.\ Sci.\
U.S.A.}, {\bf 63}, 292-299 (1969).
\bibitem{Pomeau} Pomeau, Y.\ Wetting in a corner and related questions.
{\it J.\ Colloid.\ Interf.\ Sci.\ }{\bf 113}, 5-11 (1986).
\bibitem{Hauge} Hauge, E.H.\ Macroscopic theory of wetting in
a wedge. {\it Phys.\ Rev.\ A} {\bf 46}, 4994-4998 (1992).
\bibitem{Neimark} Neimark, A.V.\ \& Kheifets, L.I.\ Solution of the problem
of the equilibrium profile of the transition zone between a wetting
film and the meniscus of the bulk phase in capillaries.
{\it Colloid J.\ of USSR} {\bf 43}, 402-407 (1981).
\bibitem{4} Rejmer, K., Dietrich, S.\ \& Napi\'{o}rkowski, M.\
Filling transition for a wedge. {\it Phys.\ Rev.\ } E {\bf 60}, 4027-4042 (1999).
\bibitem{5} Parry, A.O., Rasc\'{o}n, C.\ \& Wood, A.J.\
Critical effects at 3D wedge wetting.
{\it Phys.\ Rev.\ Lett.\ } {\bf 85}, 345-348 (2000).
\bibitem{6} Rasc\'{o}n, C.\ \& Parry, A.O.\
Geometry dependent critical exponents at complete wetting.
{\it  J.\ Chem.\ Phys.\ } {\bf 112}, 5175-5180 (2000).
\bibitem{8} Israelachvili, J.\ {\it Intermolecular \& Surface Forces},
(Academic Press, London, 1991).
\bibitem{9} Marconi, U.M.B.\ \& Van Swol, F.\ Microscopic model for hysteresis
and phase equilibria of fluids confined between parallel plates.
{\it Phys.\ Rev.\ } A {\bf 39}, 4109-4116 (1989).
\end{thebibliography}
\end{document}